\documentclass[5p,times,twocolumn]{elsarticle}
\usepackage{amsmath,amssymb}
\biboptions{comma,sort&compress}
\usepackage[utf8]{inputenc}
\usepackage{graphicx}
\usepackage{bm}
\usepackage{hyperref} 
\hypersetup{
colorlinks = true,
linkcolor = blue,
anchorcolor = blue,
citecolor = blue,
filecolor = blue,
urlcolor = blue
}
\usepackage{ulem}
\graphicspath{{Figs/}}

\renewcommand{\bar}{\overline}
\newcommand{\repi}{\text{Re}\ensuremath{^{-1}_{\pi}}}
\newcommand{\rePi}{\text{Re}\ensuremath{^{-1}_{\Pi}}}

\journal{Physics Letters B}

\begin{document}

\begin{frontmatter}

\title{Non-conformal attractor in boost-invariant plasmas}

\author[oh]{Chandrodoy Chattopadhyay}
\ead{chattopadhyay.31@osu.edu}
\author[in]{Sunil Jaiswal}
\ead{sunil.jaiswal@tifr.res.in}
\author[oh]{Lipei Du}
\ead{du.458@osu.edu}
\author[oh]{Ulrich Heinz}
\ead{heinz.9@osu.edu}
\author[in]{Subrata Pal}
\ead{spal@tifr.res.in}

\address[oh]{Department of Physics, The Ohio State University, Columbus, Ohio 43210-1117, USA}
\address[in]{Department of Nuclear and Atomic Physics, Tata Institute of Fundamental Research, Mumbai 400005, India}
\date{\today}

\begin{abstract} 
We study the dissipative evolution of (0+1)-dimensionally expanding media with Bjorken symmetry using the Boltzmann equation for massive particles in relaxation-time approximation. Breaking conformal symmetry by a mass induces a non-zero bulk viscous pressure in the medium. It is shown that even a small mass (in units of the local temperature) drastically modifies the well-known attractor for the shear Reynolds number previously observed in massless systems. For generic nonzero particle mass, neither the shear nor the bulk viscous pressure relax quickly to a non-equilibrium attractor; they approach the hydrodynamic limit only late, at small values of the inverse Reynolds numbers. Only the longitudinal pressure, which is a combination of thermal, shear and bulk viscous pressures, continues to show early approach to a far-off-equilibrium attractor, driven by the rapid longitudinal expansion at early times. Second-order dissipative hydrodynamics based on a gradient expansion around locally isotropic thermal equilibrium fails to reproduce this attractor. 
\end{abstract}

\begin{keyword}
Relativistic fluid dynamics\sep Relativistic heavy-ion collisions\sep quark-gluon plasma
\end{keyword}

\end{frontmatter}


\noindent\textit{\textbf{1.\ Introduction.}}
%
Hydrodynamics is an effective macroscopic theory that describes long wavelength excitations in a fluid. Generically it is expected to break down in systems with very large spatial or temporal gradients, reflecting significant changes in macroscopic variables over the length of one microscopic mean free path. Recent years have witnessed a paradigm shift in understanding the domain of applicability of modern formulations of relativistic dissipative hydrodynamics \cite{Heller:2011ju, Heller:2013fn,  Heller:2015dha, Kurkela:2015qoa, Blaizot:2017lht, Romatschke:2017vte, Spalinski:2017mel,  Strickland:2017kux, Romatschke:2017acs, Behtash:2017wqg, Blaizot:2017ucy, Romatschke:2017ejr, Kurkela:2018wud, Mazeliauskas:2018yef,  Behtash:2019txb, Heinz:2019dbd, Blaizot:2019scw, Blaizot:2020gql, Blaizot:2021cdv}. A major cause of this shift stems from the success of these hydrodynamic theories in describing final-state observables in ultra-relativistic collisions not only of heavy ions, but also of small nuclei such as protons  \cite{Romatschke:2007mq, Song:2007ux, Song:2008si, Schenke:2010nt, Heinz:2013th} where perhaps one does not expect the medium to locally thermalize. To better understand its unexpected effectiveness even in such extreme situations an increasing number of studies have appeared that test its performance in simplified situations where the underlying microscopic dynamics can be solved exactly. If the medium is weakly coupled, as is naively expected for a quark-gluon plasma at extremely high temperature \cite{Muller:1985, Yagi:2005yb}, the relativistic Boltzmann equation offers itself for a microscopic description. Following a long list of previous works comparing collective dynamics from the Boltzmann equation with relativistic fluid dynamics \cite{Florkowski:2013lya, Florkowski:2014sfa, Denicol:2014xca, Denicol:2014tha, Heinz:2015gka, Molnar:2016gwq, Martinez:2017ibh, Strickland:2017kux, Chattopadhyay:2018apf, Strickland:2018ayk, Jaiswal:2019cju, Kurkela:2019set, Denicol:2019lio, Almaalol:2020rnu}, we here study the solution of the Boltzmann equation in Relaxation Time Approximation (RTA) \cite{Anderson_Witting_1974} for a massive gas undergoing (0+1)-dimensional expansion with Bjorken symmetry \cite{Bjorken:1982qr}. By breaking the conformal symmetry imposed in most of the earlier works, we allow for both nonzero shear  and bulk viscous pressures, $\pi^{\eta\eta}$ and $\Pi$. We will explore the existence of an attractor in the 2-dimensional plane spanned by the normalized viscous stresses $\bar\pi=\pi^\eta_\eta/P$ and $\bar\Pi=\Pi/P$ where $P$ the thermal pressure.

The work presented here was motivated by an ongoing investigation \cite{JCDHP21, BBCJJY21} of the existence and properties of attractor solutions to the second-order dissipative fluid dynamic equations obtained by Denicol et al. \cite{Denicol:2012cn} and Jaiswal et al. \cite{Jaiswal:2014isa} for the same system. To clarify some unexpected characteristics of the solutions of the macroscopic hydrodynamic theory we had to understand the (exact) solutions of the underlying microscopic kinetic theory on which we report here. 

We work in Milne coordinates, with proper time $\tau{\,\equiv\,}\sqrt{t^2 -z^2}$ (where $t$ is time and $z$ the longitudinal Cartesian coordinate), space-time rapidity $\eta{\,\equiv\,}\tanh^{-1}(z/t)$, and metric $g_{\mu\nu} = {\rm diag}(1, -1 , -1, - \tau^2)$. In these coordinates Bjorken flow appears static, $u^\mu{\,=\,}(1, \vec{0})$, and all macroscopic quantities depend only on proper time.\\[-2ex]

\noindent\textit{\textbf{2.\ RTA Boltzmann equation for Bjorken flow.}}
%
The single particle distribution function $f(x,p)$ for a system of particles with Bjorken symmetry depends on proper time $\tau$, the transverse momentum $p_T \equiv \sqrt{(p^x)^2 + (p^y)^2}$ and the boost invariant momentum variable $w \equiv p_{\eta} \equiv t p^z - z E_p$ \cite{Florkowski:2013lya}, where $E_p = \sqrt{\bm{p}^2 + m^2}$ denotes the on-shell energy of particles with mass $m$ and 3-momentum $\bm{p}$. We here assume that it evolves according to the Boltzmann equation with a collision term in the relaxation time approximation (RTA):
\begin{equation}
\label{RTA}
	\frac{\partial f}{\partial \tau} = - \frac{f - f_{\rm eq}}{\tau_R(\tau)}.
\end{equation}
We assume Boltzmann statistics such that the local equilibrium distribution $f_{\rm eq} = \exp(-p^\tau/T)$, where $T(\tau)$ is the local temperature and $p^\tau = \sqrt{p_T^2 + w^2/\tau^2 + m^2}$ is the particle energy in the comoving frame. The relaxation time $\tau_R$ sets the timescale for equilibration and is parametrized as $\tau_R = 5 C/T$ where $C$ is a unitless constant. Eq.~(\ref{RTA}) can be converted into an integral equation for $f$ \cite{Florkowski:2013lya, Florkowski:2014sfa}:
\begin{align}
\label{BE_soln}
    f(\tau; p_T, w) =& D(\tau,\tau_0)\, f_\mathrm{in}(\tau_0; p_T,w)
\nonumber \\ 
    &+ \int_{\tau_0}^{\tau} \, \frac{d\tau'}{\tau_R(\tau')} \, D(\tau,\tau') \, f_\mathrm{eq} (\tau'; p_T, w),
\end{align} 
where $f_{\mathrm{in}}$ is the initial distribution. The damping function
\begin{equation}
    D(\tau_2,\tau_1) = \exp\left(- \int_{\tau_1}^{\tau_2} \frac{d\tau'}{\tau_R(\tau')} \right)
\end{equation}
depends on the scattering rate $1/\tau_{R}$ and controls the rate at which the distribution function loses memory of its initial form. Solving Eq.~\eqref{BE_soln} involves an iterative process \cite{Florkowski:2013lya} for obtaining the proper time evolution of the temperature by Landau matching, i.e. by identifying $T(\tau)$ as the temperature of a fictitious equilibrium state with the local comoving energy density $\epsilon(\tau)$:  
\begin{equation} 
\label{LM}
	\epsilon \equiv \langle (u \cdot p)^2 \rangle = \epsilon_{\rm eq}(T) \equiv \langle (u \cdot p)^2 \rangle_{\rm eq}\ \Leftrightarrow\ \langle (u \cdot p)^2 \rangle_{\delta} = 0.
\end{equation}
Here $\langle (\dots) \rangle{\,\equiv\,}\int_p \, (\dots) \, f$, $\langle (\dots)  \rangle_{\rm eq}{\,\equiv\,}\int_p \, (\dots) \, f_{\rm eq}$, and $\langle (\dots)  \rangle_{\delta} \equiv \int_p \, (\dots) \, ( f {-}f_{\rm eq})$ denote momentum moments of distribution functions with integration measure $\int_p \equiv \int d^2 p_T dw/ [(2\pi)^3 \tau p^\tau]$. Once the iteration for $T(\tau)$ has converged, the exact solution for $f$ is obtained from the integral equation by numerical quadrature, and the evolution of the energy-momentum tensor $T^{\mu\nu} \equiv \langle (p^\mu p^\nu) \rangle$ of the fluid can be obtained with arbitrary numerical precision. The thermal ($P$), shear ($\pi^{\mu\nu}$) and bulk ($\Pi$) viscous pressures are obtained by decomposing $T^{\mu\nu}$ as
\begin{equation}
\label{Tmunu}
	T^{\mu\nu} = \epsilon \, u^\mu u^\nu - \left(P  + \Pi\right) \Delta^{\mu\nu} + \pi^{\mu\nu} 
\end{equation}
with $\pi^{\mu\nu}u_\nu=\pi^\mu_\mu=0$. Here $\Delta^{\mu\nu} \equiv g^{\mu\nu} - u^\mu u^\nu$ is the spatial projector in the comoving frame, i.e. orthogonal to four-velocity $u^\mu$. Bjorken symmetry dictates the energy momentum tensor to be diagonal, $T^{\mu\nu} = {\rm diag} (\epsilon, P_T, P_T, P_L)$, where $P_T$ and $P_L$ are the effective transverse and longitudinal pressures, respectively. They are expressed in terms of $P$, $\Pi$, and the single independent shear stress tensor component $\pi \equiv -\tau^2 \pi^{\eta \eta}$ as $P_T = P + \Pi + \pi/2$ and $P_L = P + \Pi - \pi$. The equilibrium pressure is given by
\begin{align}
	P & \equiv - \frac{1}{3} \left\langle \left( m^2 - (u \cdot p)^2 \right) \right\rangle_{\rm eq} = - \frac{m^2}{3} \langle (1) \rangle_{\rm eq} + \epsilon/3.
\label{Pi_P2}
\end{align}
%

\noindent\textit{\textbf{3.\ Kinetic bounds on shear and bulk stresses.}}
%
As $f(x,p){\,\geq\,}0$, the effective longitudinal and transverse pressures cannot be negative in kinetic theory: $P_T{\,=\,}\frac{1}{2} \langle p_T^2 \rangle{\,\geq\,}0$, $P_L{\,=\,}\langle w^2 \rangle /\tau^2\geq0$. Also, the sum of equilibrium and bulk viscous pressures has to be non-negative: $P + \Pi \equiv \frac{1}{3} \langle  p_T^2 + (w/\tau)^2 \rangle \geq 0$. Moreover, the trace of energy momentum tensor satisfies $T_\mu^\mu \equiv m^2 \langle (1) \rangle = \epsilon -3(P{+}\Pi) \geq 0$. For the normalized stresses $\bar\pi=\pi/P$ and $\bar\Pi=\Pi/P$ we thus obtain the following four constraints:
\begin{equation}
\label{B2}
	\bar\Pi + \frac{1}{2}\bar\pi \geq -1, \quad  
	\bar\Pi - \bar\pi \geq -1, \quad 
	\bar\Pi \geq - 1, \quad 
	\bar\Pi \leq \frac{\epsilon}{3P} - 1. 
\end{equation}
Eq.~(\ref{Pi_P2}) implies that for a noninteracting massive gas the energy density always exceeds three times the thermal pressure. Eqs.~(\ref{B2}) do not depend on Bjorken symmetry and hold for arbitrary collective flow profiles whose underlying dynamics admit a kinetic description. They are also independent of the quantum statistics of the constituent particles.

\begin{figure}[t!]
\begin{center}
 \includegraphics[width=0.8\linewidth]{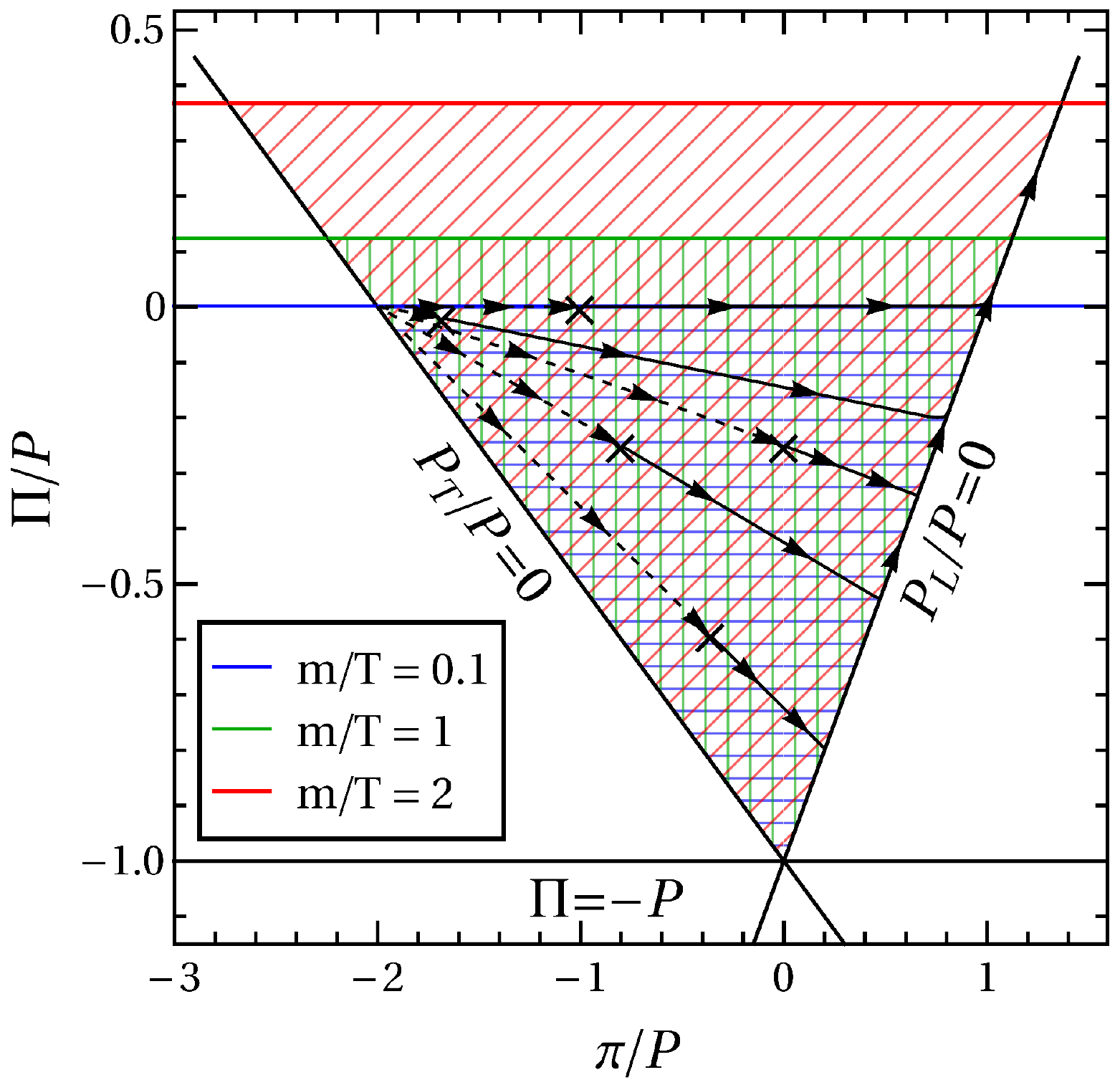}
\end{center}
\vspace{-5mm}
\caption{%
    The bounds imposed  by kinetic theory on the scaled shear and bulk viscous stresses $\bar\pi=\pi/P$, $\bar\Pi=\Pi/P$, for different choices of $m/T$. Solid black lines are free-streaming trajectories, evolving from initial states indicated by black crosses by following the arrows. Dashed black lines emerging from the same black crosses represent free-streaming evolution backward in proper time.
	\label{F1}
	\vspace*{-5mm}}
\end{figure} 

Fig.~\ref{F1} illustrates the bounds (\ref{B2}) on the scaled shear and bulk viscous stresses, together with free-streaming solutions of Eq.~(\ref{RTA}) for zero collision term, to be discussed below. The upper bound on $\Pi/P$ is a function of $\epsilon/P$ and thus depends on the particle mass through the ratio $m/T$. Solid blue, green and red lines indicate this bound for $m/T\,=\,0.1, 1, 2$, respectively. For each $m/T$ the bounds (\ref{B2}) form a triangle whose area (marked by shadings in the corresponding color) defines the shear and bulk stress pairs $(\bar\pi,\bar\Pi)$ that can be realized in kinetic theory. The allowed area grows with the particle mass $m/T$, permitting larger magnitudes for $\pi/P$ and larger positive values for $\Pi/P$ as this ratio increases. For a conformal gas of massless particles, $m{\,=\,}0$, the bulk viscous pressure $\Pi$ vanishes and the allowed region shrinks to the line $\Pi/P{\,=\,}0$. We emphasize, however, that even for an arbitrarily small non-zero mass $m/T{\,>\,}0$, $-\Pi$ can become as large as the thermal equilibrium pressure.\\[-1.5ex]

\noindent\textit{\textbf{4.\ Initial profile for large shear and bulk stresses.}}
%
To create a system with large initial bulk and/or shear stress we take an initial distribution of the form
\begin{equation}\label{f_in}
	f_{\rm in} \equiv \frac{1}{\alpha_0} \exp\left(- \frac{\sqrt{p_T^2 + (1+\xi_0) (w/\tau_0)^2 + m^2}}{\Lambda_0} \right).
\end{equation}
The parameter $\xi_0$ sets the initial anisotropy in momentum space, the slope parameter $\Lambda_0$ controls the sharpness of the distribution, and the normalisation parameter $\alpha_0$ ensures the Landau matching condition at the initial time, $\epsilon_0 = \epsilon_\mathrm{eq}(T_0)$ for a prescribed initial temperature $T_0$. Throughout this paper we impose initial conditions at $\tau_0 = 0.1$\,fm/$c$ with initial temperature $T_0 = 500$\,MeV. A sharp ($\Lambda_0 /T_0 \ll 1$) but nearly isotropic ($\xi_0 \approx 0$) initial distribution results in large negative bulk stress $(\Pi/P)_0$ and small positive shear stress $(\pi/P)_0$. A wide ($\Lambda_0 /T_0 \gg 1$) and highly anisotropic distribution reverses their relative size. Physically, substantial negative bulk stress is obtained by populating the low momentum states in phase-space with an arbitrarily large number of particles\footnote{%
    This is permissible for systems without conserved particle number.} 
(by making both $\alpha_0$ and $\Lambda_0/T_0$ small) such that the isotropic pressure $P{+}\Pi \approx 0$ and the initial energy density $\epsilon_0$ is essentially generated by the rest masses of the particles. In this sense the distribution corresponding to $\Pi/P \approx -1$ is reminiscent of a Bose condensate. This suggests that quantum effects become important near the lower corner of the allowed region in Fig.~\ref{F1}, invalidating the Boltzmann approximation. However, for the sake of comparison of kinetic theory with results from hydrodynamics using transport coefficients derived from a Boltzmann gas (as presented below), we will continue using Boltzmann statistics throughout this paper. Note that for small $z_0 \equiv m/T_0$, an upper bound on the normalised bulk that can be generated by $f_{\mathrm{in}}$ is given by $\bar\Pi_{\mathrm{max}} \approx z_0^2/6$. We take $m=50$\,MeV for the particle mass such that $m/T$ is initially small and at early times the fluid's equation of state and transport coefficients are close to their conformal limits.\\[-1.5ex]

\noindent\textit{ \textbf{5.\ Free-streaming fixed lines and fixed points.\,}}%
%
The free-streaming solution of Eq.\,(\ref{RTA}) is simply $f_\mathrm{fs}(\tau; p_T, w) = f_\mathrm{in}(p_T,w)$. Written in terms of the usual longitudinal momentum variable $p^z$ in the fluid rest frame, all free-streaming solutions, $f_\mathrm{fs}(\tau; p_T, p^z) = f_\mathrm{in}(p_T, p^z (\tau/\tau_0))$, become sharply peaked in $p^z$ as time increases. That is why the solid black lines in Fig.~\ref{F1} all approach the line of vanishing longitudinal pressure, $P_L{\,=\,}0$, eventually settling on it and following it towards larger shear stresses $\bar\pi$, driven by growing $m/T$. Thus, $P_L{\,=\,}0$
acts as an \textit{attractive fixed line} for all free streaming trajectories, characterized by a longitudinal momentum dependence $f \sim \delta(p^z)$ for the distribution function. -- The dashed black curves in Fig.~\ref{F1} are obtained by free-streaming the initial conditions {\it backward} in proper time. They are all seen to be attracted by the point $(\bar{P}_T^* = 0, \bar{\Pi}^* = 0)$, which clearly identifies it as a \textit{repulsive fixed point} for {\it forward} evolution.\footnote{%
    As $\tau \to 0$, free-streaming solutions $f_{\mathrm{fs}} = f_\mathrm{in}(p_T, p^z (\tau/\tau_0))$ become flat in $p_z$. Hence, the moments $P_T, P_L$, and $\epsilon \approx 3P$ all tend to infinity. However, the ratios $P_L/P\to3$ and $P_T/P\to0$ remain finite, leading to $ \pi/P \to -2$ and $\Pi/P \to 0$. We emphasize that the vanishing of $P_T/P$ at the repulsive fixed point \textit{does not} imply that the system approaches transverse free-streaming during backward evolution, i.e., $f_{\mathrm{fs}} (\tau \approx 0) \nsim \delta(p_T)$.}
It is also clear that any system initialized on the $P_T{\,=\,}0$ line (corresponding to initial distribution functions $f_\mathrm{in} \sim \delta(p_T)$) stays on this line, moving towards zero $P_L$ by longitudinal Bjorken expansion. However, any initial state with even infinitesimally small positive transverse pressure will move away from this line, identifying it as a repulsive fixed line. 
-- The intersection of the two fixed lines $P_L{\,=\,}0$ and $P_T{\,=\,}0$
yields the fixed point $(\bar{\Pi}^*,\bar{\pi}^*) = (-1,0)$,
where the momentum distribution takes the spherically symmetric form $f \sim \delta(|\bm{p}|)$. Physically, this represents a system where all particles are at rest such that there is no pressure but only rest mass energy. Of course, if a non-interacting system is initialised exactly at this point, it will never evolve. However, if any momentum component is only slightly nonzero, free-streaming will move the system first to the $P_L{\,=\,}0$ line (by longitudinal expansion), avoiding the point $(\bar{\Pi}^*,\bar{\pi}^*) = (-1,0)$, and then upwards on this line to $\pi\to\infty$ as $T\to0$ and $m/T\to\infty$. The only way to reach the fixed point at $(\bar{\Pi}^*,\bar{\pi}^*) = (-1,0)$ dynamically is by initializing it with $P_T=0$ exactly and letting the non-zero initial longitudinal pressure $P_L$ decay by longitudinal Bjorken expansion. Therefore this point is a saddle point -- attractive in $P_L$ direction but repulsive in $P_T$ direction. 

Note that for $m{\,\ne\,}0$ the free-streaming system has no attractive fixed point at all: all initial conditions eventually hit the $P_L=0$ line (``late time free-streaming attractor'') and then follow it to $(\bar\pi,\bar\Pi)=(\infty,\infty)$ as the ratio $m/T$ grows beyond all bounds. The system never thermalizes.   

The free-streaming solution described in this section dominates the RTA Boltzmann solution (\ref{BE_soln}) for large Knudsen number, i.e. for $\tau{\,\ll\,}\tau_R(\tau){\,=\,} 5C/T(\tau)$. For times $\tau{\,\gg\,}\tau_R(\tau)$ the Knudsen number is small and the system thermalizes, following the Navier-Stokes attractor. In conformal Bjorken systems the early-time free-streaming dynamics joins smoothly the late-time Navier-Stokes dynamics along a ``far-off-equilibrium attractor'' which some versions of fluid dynamics describe with excellent precision \cite{Heller:2011ju, Heller:2013fn,  Heller:2015dha, Kurkela:2015qoa, Blaizot:2017lht, Romatschke:2017vte, Spalinski:2017mel, Strickland:2017kux, Romatschke:2017acs, Behtash:2017wqg, Blaizot:2017ucy, Romatschke:2017ejr, Kurkela:2018wud, Mazeliauskas:2018yef, Behtash:2019txb, Heinz:2019dbd, Blaizot:2019scw, Blaizot:2020gql, Blaizot:2021cdv, Chattopadhyay:2018apf, Strickland:2018ayk, Jaiswal:2019cju, Kurkela:2019set, Denicol:2019lio, Almaalol:2020rnu}. To explore whether this continues to hold true for non-conformal systems we now contrast the results of this section with the early-time attractor structure of second-order non-conformal viscous hydrodynamics.\\[-1.5ex]

\noindent\textit{\textbf{6.\ Second-order non-conformal hydrodynamics.}}
%
For a massive gas undergoing Bjorken expansion, the second-order hydrodynamic evolution equations for the energy density and the bulk and shear viscous stresses read \cite{Denicol:2014vaa, Jaiswal:2014isa}
\begin{align}
    \frac{d\epsilon}{d\tau} &= -\frac{1}{\tau}\left(\epsilon + P + \Pi -\pi\right) \, ,  \label{epsBj}
\\
    \frac{d\Pi}{d\tau} + \frac{\Pi}{\tau_\Pi} &= -\frac{\beta_\Pi}{\tau} - \delta_{\Pi\Pi}\frac{\Pi}{\tau} +\lambda_{\Pi\pi}\frac{\pi}{\tau} \, ,  
\label{bulkBj}
\\
    \frac{d\pi}{d\tau} + \frac{\pi}{\tau_\pi} &= \frac{4}{3}\frac{\beta_\pi}{\tau} - \left( \frac{1}{3}\tau_{\pi\pi} + \delta_{\pi\pi} \right) \frac{\pi}{\tau} + \frac{2}{3} \lambda_{\pi\Pi} \frac{\Pi}{\tau} \,.
\label{shearBj}
\end{align}
The transport coefficients $\beta_\Pi$, $\delta_{\Pi\Pi}$, $\lambda_{\Pi\pi}$, $\beta_\pi$, $\delta_{\pi\pi}$, $\tau_{\pi\pi}$ and $\lambda_{\pi\Pi}$ are calculated from the RTA Boltzmann equation in second-order Chapman-Enskog approximation \cite{Jaiswal:2014isa}; they are functions of
$m/T$ times a power of $T$ corresponding to their dimension. 

To compare the behavior of the solutions to these equations at very early times with those in Sec.\,5 we study the limit of large Knudsen numbers $\tau{\,\ll\,}\tau_R$, corresponding to the free-streaming limit.\footnote{%
    Eqs.~(\ref{epsBj})-(\ref{shearBj}) were derived by expanding in powers of (small) Knudsen number; we study them here for large Knudsen number to see how they fail.}
In this limit the temperature is high and we can use transport coefficients evaluated in the $m{\,=\,}0$ limit: $\frac{\beta_\Pi}{\epsilon+P}{\,=\,}0$,  $\delta_{\Pi\Pi}{\,=\,}\frac{2}{3}$, $\lambda_{\Pi\pi}{\,=\,}0$, $\frac{\beta_\pi}{\epsilon+P}{\,=\,}\frac{1}{5}$,   $\delta_{\pi\pi}{\,=\,}\frac{4}{3}$,  $\tau_{\pi\pi}{\,=\,}\frac{10}{7}$, and $\lambda_{\pi\Pi}{\,=\,}\frac{6}{5}$. Under these approximations second-order non-conformal hydrodynamics yields the following three fixed points for the pressure-normalized viscous stresses:
\begin{align}
\label{fp:non_conf_2}
    \bigl\{\bar{\Pi}^*, \bar{\pi}^*\bigr\} 
    = \{0,1.214\};\  \{0,-2.64\};\  \{-3.56,-1.56\}.
\end{align}
The first two agree with corresponding fixed points of second-order \textit{conformal} hydrodynamics where $\Pi{\,\equiv\,}0$; the third one arises from non-conformality. These three fixed points may be viewed as crude (hydrodynamic) approximations\footnote{%
    Note that all three violate the bounds imposed by kinetic theory!} 
of the three corners of the blue $(m/T{\,\approx\,}0)$  triangular region in Fig.\,\ref{F1}. For an in-depth analysis of their nature we refer to Ref.~\cite{JCDHP21}.\\[-1.5ex]

\noindent\textit{\textbf{7.\ Dynamics at finite Knudsen number.}}
%
We now discuss the evolution of the shear and bulk viscous stresses for finite relaxation time $\tau_R(\tau)= 5C/T(\tau)$, by comparing the exact solution of the RTA Boltzmann equation (\ref{BE_soln}) with its hydrodynamic approximation (\ref{epsBj})-(\ref{shearBj}). The discussion of the free-streaming limit in Secs.~5 and 6 will help to understand the early-time behaviour in the micro- and macroscopic approaches, as well as their differences.\\[-2ex]

\noindent\textit{\textbf{a.\ Dynamics with large initial bulk viscous pressure.}}
%
\begin{figure}[t!]
\centering
\includegraphics[width=\linewidth]{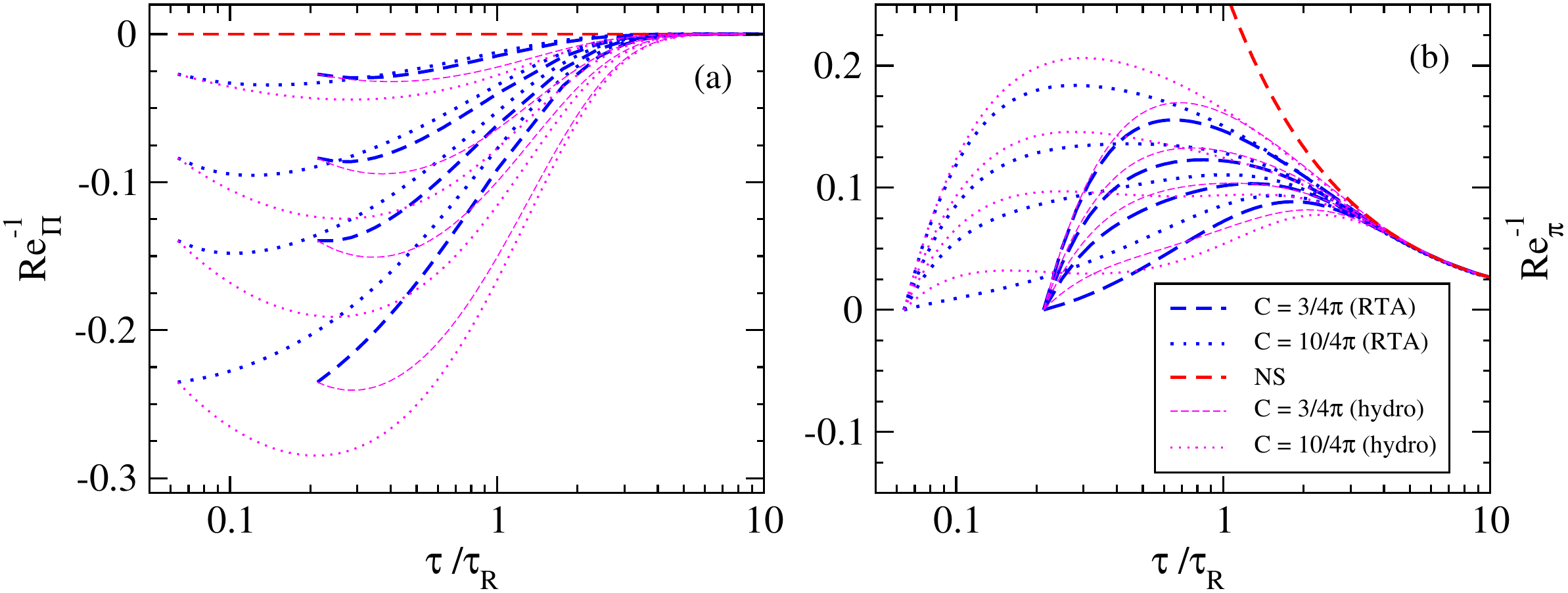}
\vspace{-5mm}
\caption{(Color online) 
    Scaled time evolution of the (a) bulk and (b) shear inverse Reynolds numbers, for isotropic initial conditions and two different choices of the relaxation parameter $C$. Blue lines are solutions of the RTA Boltzmann equation; magenta ones are obtained with second-order Chapman-Enskog hydrodynamics. NS indicates the Navier-Stokes limit. 
    \label{F2}}
\end{figure} 
%
Figure~\ref{F2} shows the evolution of the bulk and shear inverse Reynolds numbers, Re$^{-1}_{\Pi} \equiv \Pi/(\epsilon{+}P)$ and  Re$^{-1}_{\pi} \equiv \pi/(\epsilon{+}P)$, as functions of the scaled time $\bar{\tau} \equiv \tau/\tau_R$. To explore the strength of the coupling between the bulk and the shear channels and the resulting modification of the well-known attractor behavior exhibited by $\repi$ in conformal fluids, we first study momentum-isotropic initial conditions with vanishing shear stress ($(\xi_0{\,=\,}0{\,=\,}\pi_0$) but large negative bulk viscous pressure. For the lowest blue curves in Fig.~\ref{F2}a we tuned $\Lambda_0$ and $\alpha_0$ to generate $\rePi{\,\approx\,}-0.23$, close to its lower limit Re$^{-1}_{\Pi}{\,\geq\,}{-}0.25$\ ($\bar\Pi{\,\geq\,}{-}1$). Different line styles correspond to different relaxation times $\tau_R = 5 C/T$, with $C = 3/(4\pi)$ (dashed) for a more strongly coupled and $C = 10/(4\pi)$ (dotted) for a more weakly coupled fluid. Here and in all subsequent figures blue curves represent solutions of kinetic theory while magenta curves (shown for comparison) are the predictions of second-order hydrodynamics, Eqs.~(\ref{epsBj})-(\ref{shearBj}), for identical initial conditions. The somewhat thicker red dashed lines show the corresponding first-order hydrodynamic (Navier-Stokes) solutions, given by (Re$^{-1}_{\Pi})_{_\mathrm{NS}} = -(\zeta/s)/(\tau T)$ in Fig.~\ref{F2}a and (Re$^{-1}_{\pi})_{_\mathrm{NS}} = \frac{4}{3}(\eta/s)/(\tau T)$ in Fig.~\ref{F2}b, respectively. The specific shear and bulk viscosities are $\eta/s = 5C \, {\cal I}/ \big((\epsilon{+}P) T \big)$,  $\zeta/s=5C \big[ {5\cal I} / \big(3(\epsilon{+}P)T\big) - c_s^2 \big]$, where ${\cal I} \equiv (1/15) \bigl\langle (p^\mu\Delta_{\mu\nu} p^\nu)^2/(u \cdot p) \bigr\rangle_{\rm eq}$ and the squared speed of sound $c_s^2 \equiv (\epsilon + P)/ \big[3(\epsilon + P) + (m/T)^2 P \big]$ \cite{Denicol:2014vaa, Jaiswal:2014isa}. Note that for $m\ne 0$ the specific viscosities are both temperature dependent, albeit weakly so as long as $m/T$ is small.\footnote{%
    $\zeta/s$ vanishes in the limit $m/T\to0$.}
Since our initial $m/T_0$ is small, the Navier-Stokes value for Re$^{-1}_{\Pi}$ shown in Fig.~\ref{F2}a (and also the one shown in Fig.~\ref{F3}a below) is close to zero throughout the system's evolution. 

The solution with the largest $|(\rePi)_0|$ corresponds to an initial momentum distribution where only states around $|\bm{p}| \approx 0$ are populated, thus it can hardly be shrunk any further by longitudinal expansion. As a result its evolution is dominated by thermalizing dynamics which broadens its momentum distribution from the beginning. Accordingly, in Fig. \ref{F2}a the kinetic theory solutions with the largest initial $|(\rePi)_0|$ exhibit monotonic growth towards equilibrium. In contrast, distributions corresponding to initial states with $|(\rePi)_0| \lesssim 0.2$ first shrink in longitudinal momentum owing to the rapid longitudinal expansion before thermalizing dynamics takes over. As a result, these blue curves decrease slightly before turning to approach the late-time NS limit. One notes that the hydrodynamic curves (in magenta) do not describe the kinetic theory trajectories very well and initially decrease for {\it all} initial conditions. For large negative initial values for Re$^{-1}_{\Pi}$ they even drop below the physically allowed kinetic theory limit Re$^{-1}_{\Pi}{\,\geq\,}{-}0.25$, leading to negative total longitudinal and transverse pressures.    

\begin{figure}[t!]
\centering
\includegraphics[width=1.01\linewidth]{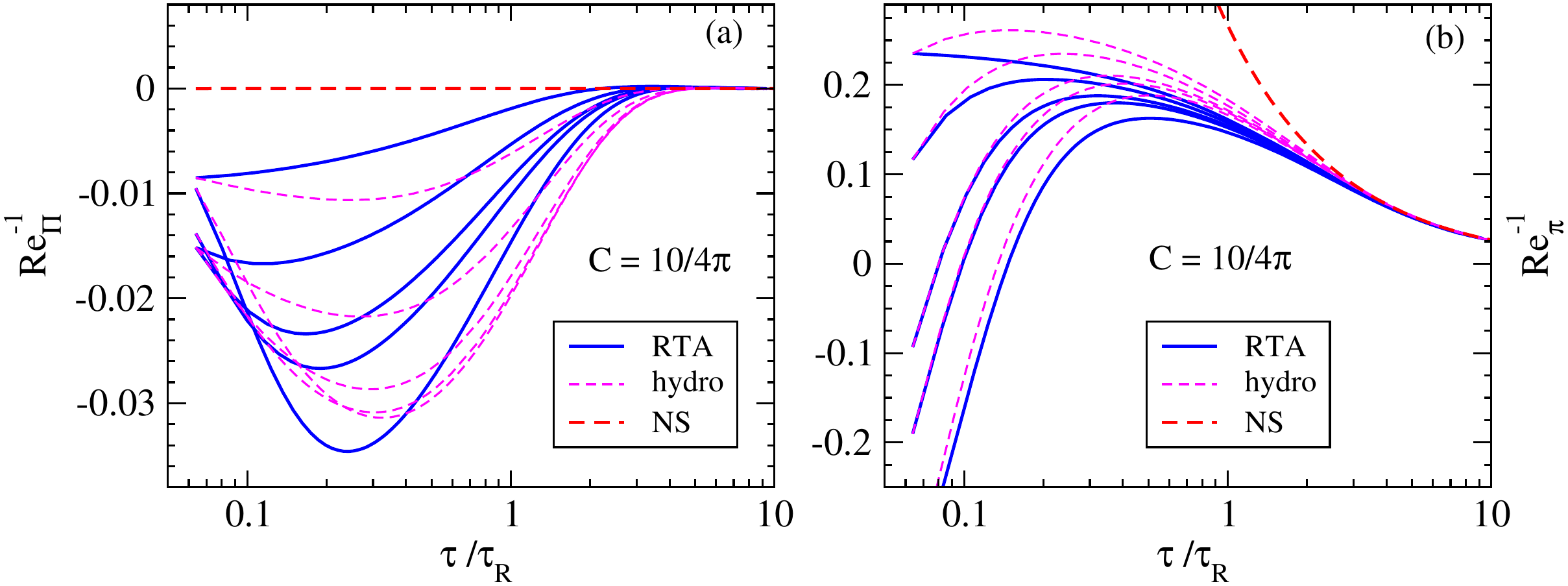}
\vspace{-5mm}
\caption{(Color online) 
    Same as in Fig.~\ref{F2} but for non-zero initial shear stress $(\repi)_0$. Solid blue lines (RTA): kinetic theory; dashed magenta lines: hydrodynamics. All solutions use $C = 10/4\pi$. 
	\label{F3}}
\end{figure} 

Turning to the shear stress in Fig.~\ref{F2}b, one sees the familiar pattern that strong longitudinal expansion renders the initially isotropic momentum anisotropic, leading to a shear inverse Reynolds number that initially increases until thermalizing processes drive it back down. The maximum shear stress developed by the system before the Bjorken expansion rate falls below the microscopic relaxation rate is smaller for the strongly coupled fluid. It also decreases with the magnitude of the initial bulk viscous pressure. This is because for initial distributions that are already sharply peaked around $|\bm{p}|{\,\approx\,}0$ it is more difficult for the longitudinal expansion to generate sizeable momentum space anisotropies. Also, Fig.~\ref{F1} shows that solutions starting from the lower part of the allowed triangular region can only generate comparatively small shear stresses during the early-time free-streaming dynamics (which moves the system towards the $P_L{\,=\,}0$ line) before collision-induced momentum isotropization drives the system again away from that line \cite{JCDHP21}. -- An important feature of Fig.~\ref{F2}b is that shear stress trajectories with different initial bulk viscous pressures \textit{repel} each other initially and collapse on the late-time hydrodynamic Navier-Stokes attractor only at $\bar\tau{\,\gtrsim\,}3$. This contrasts strongly with the pattern observed in conformal systems where (in the absence of bulk viscosity) trajectories with different initial shear stresses rapidly approach an early-time attractor on a much shorter time scale controlled by the initialization time $\tau_0$. This difference will be studied more in Fig.~\ref{F3}. Finally, we note that the hydrodynamic curves (magenta), while sharing this absence of an early-time attractor, do not agree well with the kinetic theory solutions, especially in the bulk sector.\\[-2ex]

\noindent\textit{\textbf{b.\ Dynamics with small bulk viscous pressure.}} 
%
For small bulk viscous pressures one naively expects to recover the known attractor structure of conformal systems. Fig.~\ref{F1} shows that, for small initial $m/T_0$, selecting initial conditions with large magnitude of the shear stress forces small initial bulk viscous pressures. Some evolution trajectories for such initial conditions are shown as solid blue lines in Fig.~\ref{F3} for $C{\,=\,}10/4\pi$. (Please note the magnified vertical scale in the left panel!). For comparison, the magenta curves are the corresponding trajectories from second-order hydrodynamics. Panel (a) shows that the bulk stress evolution is qualitatively similar to the initially momentum-isotropic case, except for the much smaller magnitudes of $|(\rePi)_0|$. There is no evidence of an early-time attractor -- convergence with the late-time NS attractor does not occur until $\bar\tau{\,\gtrsim\,}3$. Panel (b) shows the same for the shear inverse Reynolds number: Even though the bulk inverse Reynolds number never exceeds a few percent, the early time attractor found for the shear stress in conformal systems is destroyed by bulk-shear coupling in Eqs.~(\ref{epsBj})-(\ref{shearBj}).\footnote{%
    For an in-depth mathematical analysis of the early-time behavior of the trajectories shown in Figs.~\ref{F2} and \ref{F3} see \cite{JCDHP21}.}    
Again, convergence with the late-time attractor is delayed until $\bar\tau{\,\gtrsim\,}3$, and at early times second-order viscous hydrodynamics really does not provide a very good approximation of the underlying kinetic theory.\footnote{%
    In Ref.~\cite{JCDHP21} it is shown that a modified version of anisotropic hydrodynamics provides much better agreement with kinetic theory.}
Figs.~\ref{F2} and \ref{F3} together establish that it is the magnitude of $\rePi$, and not of $m/T$, which plays the dominant role in controlling the system's deviation from conformality.\\[-2ex]

\noindent\textit{\textbf{c.\ Early-time attractor.}}
%
With neither shear nor bulk stress evolution controlled by an early-time, far-off-equilibrium attractor, is there any such attractor at all in non-conformal systems? The answer is: Yes. The key to finding it is the realization in Sec.~5 above that in Bjorken flow the RTA Boltzmann equation is approximately free-streaming at early times, and that the line $P_L{\,=\,}0$ acts as an attractor for this approximate free-streaming dynamics \cite{Jaiswal:2019cju, Kurkela:2019set}. Whereas in conformal systems undergoing Bjorken flow the longitudinal pressure $P_L$ (or rather its deviation from the thermal pressure $P(\epsilon)$) and the shear stress $\pi$ are equivalent physical quantities and mutually interchangeable (which means that $P_L$ and $\pi$ share a common attractor), this is no longer true in non-conformal systems. We will now show that for the RTA Boltzmann equation with Bjorken flow the existence of the attractive ``fixed line" $P_L{\,=\,}0$ for free-streaming dynamics entails for the full solution with non-zero collision term an early-time attractor for the longitudinal pressure $P_L$, but that this does not also imply the existence of such attractors for the shear and bulk viscous stresses. In other words, only $P_L{\,=\,}P{+}\Pi{-}\pi$ (or, equivalently, $\Pi{-}\pi$) has an early-time attractor, and it is driven by the approximately free-streaming dynamics at early times of the RTA Boltzmann equation in Bjorken flow. As a corollary, similar early-time attractors are not expected in other systems in which the early-time dynamics is not dominated by free-streaming.

%
\begin{figure}[t!]
\centering
 \includegraphics[width=0.8\linewidth]{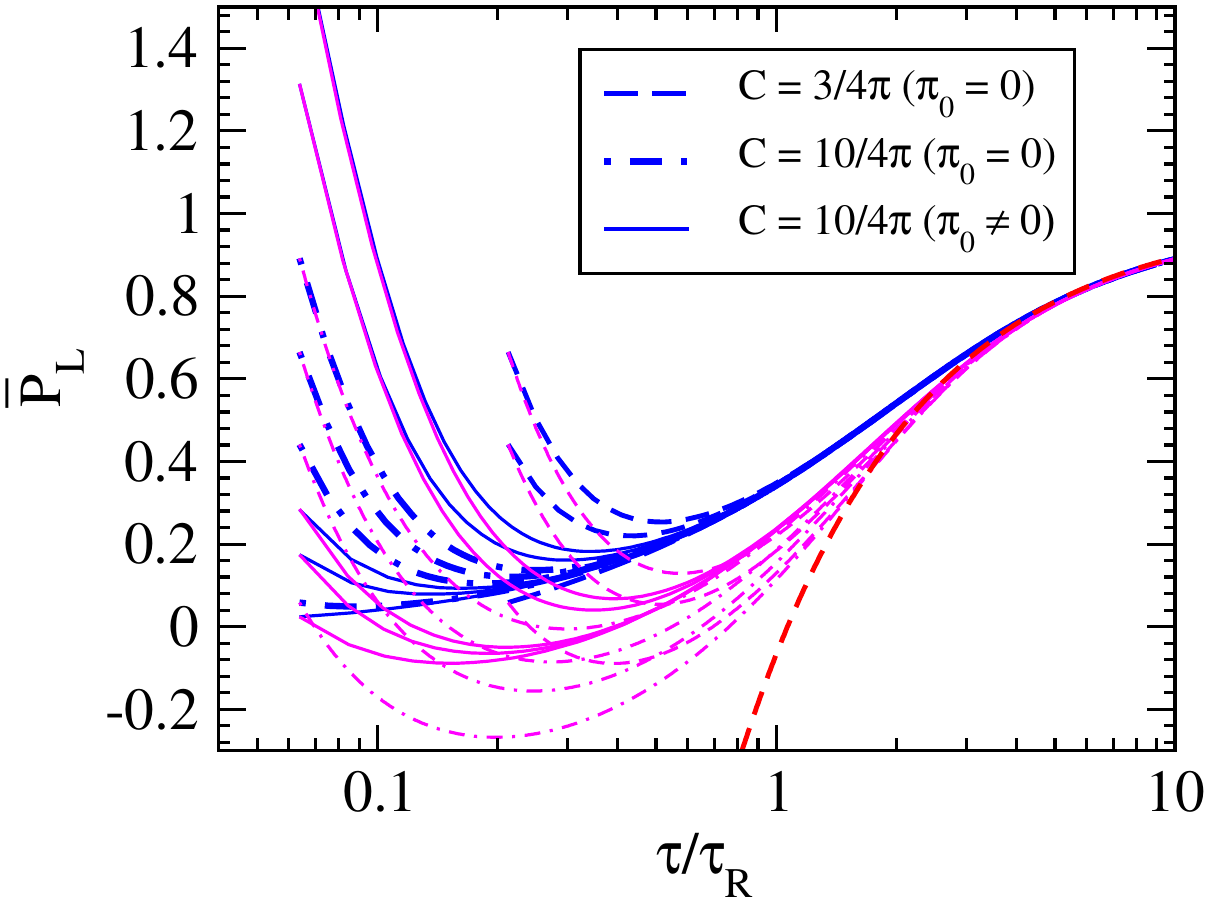}
 \vspace{-3mm}
 \caption{(Color online) 
 Evolution of the scaled longitudinal pressure obtained from the RTA Boltzmann equation with $C = 3/4\pi$ for vanishing initial $(\repi)_0$ (blue dashed lines), and with $C = 10/4\pi$ for vanishing (blue dashed-dotted lines) and non-vanishing $(\repi)_0$ (blue solid lines). The corresponding magenta curves are obtained using second-order hydrodynamics. The thicker red dashed line is the first-order NS solution with $C = 10/4\pi$.
	\label{F4}}
	 \vspace{-2mm}
\end{figure} 
%

Figure~\ref{F4} shows the evolution of the scaled longitudinal pressure $\bar{P}_L=P_L/P$ from the RTA Boltzmann equation as a function of the scaled time $\bar\tau = \tau/\tau_R$.\footnote{%
    A slightly different quantity, $A_1 \equiv -1 + \repi - \rePi$ (note the different normalizing factor) was previously explored in Ref.~\cite{Romatschke:2017acs}. Its fixed point structure at early times is non-universal. It was shown in \cite{Romatschke:2017acs} that $A_1$ shows late-time attractor behavior when plotted versus the `gradient strength', $\Gamma = \tau/\gamma_s$ where $\gamma_s = [(4/3)\eta + \zeta]/(\epsilon+P)$.}
The blue dashed and dash-dotted lines correspond to isotropic initial conditions using $C = 3/(4\pi)$ and $C = 10/(4\pi)$, respectively, whereas the blue solid lines are obtained using $C = 10/4\pi$ with anisotropic initial distributions.\footnote{%
    These initial conditions are identical to those used in Figs.~\ref{F2}, \ref{F3}.}
The corresponding solutions from second-order hydrodynamics are plotted in magenta for comparison. All kinetic theory solutions are seen to join already at early times $\bar\tau{\,\lesssim\,}0.5$ a universal attractor that starts from $P_L/P{\,\approx\,}0$ at $\bar\tau_0{\,\to\,}0$. As the system begins to isotropize this universal curve approaches unity, joining the first-order hydrodynamic NS attractor at $\bar\tau{\,\gtrsim\,}4$. The hydrodynamic trajectories, on the other hand, do not exhibit a universal early-time attractor; universality is only seen after they join the NS attractor at $\bar\tau{\,\gtrsim\,}4$. Clearly, second-order hydrodynamics is not a very accurate approximation of the underlying kinetic theory when $\bar\tau{\,<\,}3$.\\[-1.5ex]

\textit{\textbf{8.\ Summary.}}
%
We studied the evolution of dissipative flows for a non-conformal (0+1)-dimensional expanding system whose microscopic dynamics is governed by the RTA Boltzmann equation. It was shown that the introduction of even a small mass (in units of the temperature) can generate large bulk viscous pressure and substantially affect the shear stress evolution via bulk-shear coupling. No attractor behavior is observed in the normalised bulk stress channel. Moreover, the well-known universal early-time attractor for the normalised shear stress observed in conformal systems is disrupted once $\Pi{\,\ne\,}0$. Only the combination $(P{+}\Pi{-}\pi)/P$, i.e. the normalized longitudinal pressure $\bar{P}_L{\,=\,}P_L/P$, continues to exhibit universal early-time attractive behavior, driven by the strong longitudinal flow at early times which renders the RTA Boltzmann dynamics effectively free-streaming. The attractor solution for $P_L/P$ smoothly joins early-time longitudinal free-streaming to late-time equilibrating dynamics, and initial deviations from this attractor relax to it via power-law decay, driven by the rapid longitudinal expansion. Second-order non-conformal hydrodynamics was found to be unable to describe this early-time attractor behavior of the underlying RTA Boltzmann kinetics; however, a modified version of anisotropic hydrodynamics to be reported in \cite{JCDHP21} avoids this failure for systems undergoing Bjorken flow.\\[-1.5ex]

\noindent \textit{\textbf{Acknowledgements:}}
The authors thank Derek Everett, Kevin Ingles, Dananjaya Liyanage and Mike McNelis for insightful comments. SJ acknowledges helpful discussions with the authors of \cite{BBCJJY21}. This work was supported by the U.S. Department of Energy, Office of Science, Office for Nuclear Physics under Award No. \rm{DE-SC0004286} and by the Department of Atomic Energy (Government of India) under Project Identification No. RTI 4002.

\bibliographystyle{elsarticle-num}
\bibliography{references}

\end{document}